\documentclass[pra,amsmath,a4paper,12pt]{revtex4}

\newcommand{\CR}{\nonumber\\ }
\renewcommand{\d}{\mathrm{d}}
\renewcommand{\i}{\mathrm{i}}
\newcommand{\fl}{}

\begin{document}

\title{Perturbation theory for sextic doubly anharmonic oscillator}
\author{I. V. \surname{Dobrovolska}}
\email{dobrovolska@resonance.zp.ua}
\author{R. S. \surname{Tutik}}
\email{tutik@ff.dsu.dp.ua}
\affiliation{Department of Physics,
Dniepropetrovsk National University, Dniepropetrovsk, 49050, Ukraine}

\begin{abstract}
A simple method for the calculation of higher orders of
the logarithmic perturbation theory for bound states of the
spherical anharmonic oscillator is developed. The structure of the
perturbation series for energy eigenvalues of the sextic doubly
anharmonic oscillator is investigated. The recursion technique
for deriving renormalized perturbation expansions is offered.
\end{abstract}

\maketitle

\section{Introduction}

The problem of the quantum anharmonic oscillator has been the
subject of much discussion for decades because of its importance
in quant-field theory, molecular physics, and solid-state and
statistical physics. However, while very extensive literature has
been devoted to the one-dimensional consideration, relatively less
attention has been given to physically more interesting but more complicated
three-dimensional case. This applies to the doubly anharmonic 
oscillator of the type
\begin{equation}
V(r)=\frac{1}{2}\omega^{2}r^{2}+\lambda r^{4}+\mu
r^6,\;\;\;\;\;\mu>0 \label{0}
\end{equation}
as well. In the one-dimensional space, there has been some interest 
in this theory both in its own right \cite{b1, b2}, and in
the corresponding scalar field theory in 2+1 dimensions
 \cite{b3, b4}. Besides double perturbation expansions in
terms of the two coupling constants $\lambda$ and $\mu$ \cite{b4,
b5}, this problem has been studied using the theory of continued
fraction \cite{b6}, too.

In the three-dimensional space, only exact solutions for the case
when the parameters satisfy certain relations have been obtained
by a number of authors \cite{b7, b8, b9, b10, b11} and first four
eigenvalues have been calculated with the Hill determinant
approach \cite{ b12}.

The main tool for studying energy eigenvalues and eigenfunctions
of the bound-state problem within the framework of the
Schr\"odinger equation is the logarithmic perturbation theory 
(for references see \cite{b5, b13}). But the nonconvergence
of obtained expansions compels us to resort to modern procedures
of summation of divergent series, such as various versions 
of the renormalization technique \cite{b5}. However, to provide 
a reasonable accuracy these procedures need to know high orders 
of the perturbation series. Unfortunately, with the standard 
approach to the logarithmic perturbation theory we can easily
obtain the higher order corrections only for the ground states,
because the description of radially excited states becomes 
extremely cumbersome.

Recently, a new procedure based on the specific quantization
conditions has been proposed to get series of the logarithmic
perturbation theory via the $\hbar$-expansions within the
framework of the one-dimensional Schr\"odinger equation
\cite{b13}. Avoiding disadvantage of the standard approach, this
straightforward semiclassical procedure results in the new handy
recursion formulae with the same simple form both for ground and
excited states. Moreover, these formulae can be easily applied to
any renormalization scheme of improving the convergence of
expansions \cite{b5}.

The object of this paper is to extend the above mentioned
formalism to the bound-state problem for the spherical anharmonic
oscillator with its subsequent applying to the doubly anharmonic oscillator.

\section{The Method}

At the beginning we shall concentrate our attention on the
bound-state problem for a non-relativistic particle moving in a
central potential of anharmonic oscillator of a general form
admitted bounded eigenfunctions and having in consequence a
discrete energy spectrum. This potential has a single simple
minimum at the origin and is given by the symmetric function
\begin{equation}
V(r)=\frac{1}{2}m\omega^{2}r^{2}+\sum_{i{\geq}1}v_{i}r^{2i+2}.
\label{2}
\end{equation}
Then after separating the angular part, the reduced radial part of
the Schr\"{o}dinger equation takes the form
\begin{equation}
 -{\frac{\hbar^2}{2m}}U''(r)+\left({\frac{{\hbar}^2l(l+1)}{2mr^2}+V(r)}\right)U(r)=EU(r).
\label{1}
\end{equation}

 With applying the substitution, $ C(r)=
\hbar U'(r) / U(r) $, accepted in the logarithmic perturbation
theory, we go over from the Schr\"{o}dinger equation (\ref{1}) to
the Riccati one
\begin{equation}
\hbar C'(r)+C^{2}(r)=\frac{{\hbar}^2l(l+1)}{r^2}+2mV(r)-2m E.
\label{3}
\end{equation}

We attempt to solve it in a semiclassical manner with series
expansions in the Planck constant
\begin{equation}
E=\sum^{\infty}_{k=0}{E_{k}\hbar^{k}},\;\;\;\;\;
C(r)=\sum^{\infty}_{k=0}C_{k}(r)\hbar^{k}. \label{4}
\end{equation}

In the complex plane, the number of zeros $N$ of a wave function
inside the closed contour is defined by its logarithmic
derivative, $C(r)$, through the relation
\begin{equation}
\frac{1}{2\pi\i}\oint{C(r)\,\d{r}}=
\frac{1}{2\pi\i}\sum^{\infty}_{k=0}{\hbar^{k}\oint{C_{k}(r )\,\d
r}}=\hbar N. \label{5}
\end{equation}

This quantization condition is exact and is widely used for deriving
higher-order corrections to the WKB-approach \cite{b14,b15} and the
$1/N$-expansions \cite{b16,b17,b18}. There is, however, one important
point to note. Because the radial and orbital quantum numbers, $n$ and $l$,
correspondingly, are specific quantum notions, 
the quantization condition (\ref{5}) must be supplemented with a rule
of achieving a classical limit for these quantities. It is this rule that stipulates the kind of the
semiclassical approximation.

In particular, within the framework of the WKB-method the
passage to the classical limit is implemented using the rule
\begin{equation}
\hbar\to 0,\;\;\;\;n \to\infty,\;\;\;\; l\to\infty,\;\;\;\;\hbar
n={\rm const},\;\;\;\;\hbar l={\rm const},\label{6}
\end{equation}
whereas the $1/N$-expansion requires the condition
\cite{b16,b17,b18}
\begin{equation}
\hbar\to 0,\;\;\;\; n={\rm const},\;\;\;\; l\to\infty,\;\;\;\;\hbar
n\to 0,\;\;\;\;\hbar l={\rm const}.\label{7}
\end{equation}

The semiclassical treatment of the logarithmic perturbation theory
proved to involve the alternative possibility:
\begin{equation}
\hbar\to 0,\;\;\;\; n={\rm const},\;\;\;\;l={\rm const},\;\;\;\;
\hbar n\to 0,\;\;\;\;\hbar l\to 0. \label{8}
\end{equation}

Let us consider the latter rule from  the physical point of view. Since
$\hbar l \rightarrow 0$ as $\hbar \rightarrow 0$, the centrifugal term,
$\hbar^2 l \left(l+1\right)/r^2$, has the second order in $\hbar$ and
disappears in the classical limit that corresponds to falling a particle
into the center. This means that a particle drops into the bottom of
a potential well as $\hbar \rightarrow 0$ and its classical energy
becomes $E_0 = \min V(r) = 0$. It appears from this that the series
expansion in the Planck constant for the energy eigenvalues must now
read as $E =
\sum_{k=1}^{\infty}{E_k\hbar^k}$.

Upon inserting the $\hbar$-expansions for $E$ and $C(r)$ into the Riccati
equation (\ref{3}) and collecting coefficients of equal powers of
$\hbar$, we obtain the following hierarchy of equations:
\begin{eqnarray}
C_{0}^{2}(r)=2 \; m V(r),\nonumber\\
C'_{0}(r)+2 \; C_{0}(r)C_{1}(r)=-2 \; m E_{1},\nonumber\\
C'_{1}(r)+2 \; C_{0}(r)C_{2}(r)+C_{1}^{2}(r)=\frac{l(l+1)}{r^2}-2 \; m E_{2},\label{9}\\
\cdots \nonumber\\
C'_{k-1}(r)+\sum_{i=0}^{k}C_{i}(r)C_{k-i}(r)=-2 \; m
E_{k},\;\;\;k>2. \nonumber
\end{eqnarray}

In the case of ground states, the recurrence system at hand
coincides with one derived by means of the standard technique and
can be solved straightforwardly. For excited states, however, it
is necessary to take into account the nodes of the wave function,
that we intend to do by making use of the quantization condition
(\ref{5}).

It should be stressed that our approach is quite distinguished
from the WKB method not only in the rule of achieving a classical
limit but also in the choice of a contour of integration in
complex plane. With a view to elucidating the last difference
let us now sketch out the WKB treatment of the bound-state problem
for the spherical anharmonic oscillator. 
In the complex plane, because the potential is described by the
symmetric function (\ref{2}), this problem has two pairs of turning
points, i.e. zeros of the classical momentum. Therefore we have two cuts
between these points: in the region $r>0$ as well as in the region $r<0$. 
In spite of only one cut lies in the physical region
$r>0$, the contour of integration in the WKB quantization condition has
to encircle both cuts for the correct result for harmonic oscillator to
be obtained \cite{b19}.

In our approach, when a particle is dropping into the bottom of a
potential well, these four turning points are drawing nearer and,
at last, are joining all together at the origin. Hence, all
nodes of the wave function are now removed from both positive and
negative sides of the real axis into the origin and our contour of
integration must enclose only this point and no other
singularities. Further, let us count the multiplicity of zero
formed at $r = 0$. For the regular solution of the equation
(\ref{1}), the behaviour $r^{l+1}$ as $r \rightarrow 0 $ brings
the value $l+1$. The number of nodes of eigenfunction in the
region $ r > 0$ is equal to radial quantum number $n$. But
because the potential (\ref{2}) is a symmetric function the same
number of zeros must be in the region $r<0$, too. Then the total
number of zeros inside the contour becomes equal to $N=2n+l+1$. 

Taking into account the first order in $\hbar$ of the right-hand
side, the quantization condition (\ref{5}) is now rewritten as:

\begin{equation}
\frac{1}{2\pi\i}\oint{C_1(r)\, \d r}=2n+l+1,\;\;\;\;\;\;\;\;
\frac{1}{2\pi\i}\oint{C_{k}(r )\,\d r}=0 ,\quad \;k\neq1.
\label{10}
\end{equation}

A subsequent application of the theorem of residues to the
explicit form of functions $C_k(r)$ easily solves the problem of
description of the radially excited states.
With that end in view let us consider the system (\ref{9}) and
investigate the behaviour of the function $C_k(r)$. From the first
equation it is seen that $C_0(r)$ can be written in the form
\begin{equation} \fl
C_0(r) = - \left[2 \, m\,  V(r) \right]^{1/2} = -m\, \omega\, r
\left( 1 + \frac{2}{m\, \omega^2} \sum_{i=1}^{\infty}{v_i
\,r^{2i}}\right)^{1/2} =\, r \sum_{i=0}^{\infty}{C_i^0 \,r^{2i}},
\label{11}
\end{equation}
where the minus sign is chosen from the boundary conditions and
coefficients $C_i^0$ are defined by parameters of the potential
through the relations:
\begin{equation}
C^{0}_{0}=-m\omega,\;\;\;\;\;\;\;\; C^{0}_{i}={\frac1{2m\omega}}
\left({\sum_{p=1}^{i-1}{C^{0}_{p} C^{0}_{i-p}-2 m v_{i}}}\right),
\;i\geq 1.\label{12}
\end{equation}

From (\ref{11}) we recognize that $C_0(0)=0$ and, consequently,
the function $C_1(r)$ has a simple pole at the origin, while
$C_k(r)$ has a pole of the order of $\left(2k-1\right)$. Hence 
it appears that
$C_k(r)$ can be represented by the Laurent series:
\begin{equation}
C_{k}(r)= r^{1-
2k}\sum^{\infty}_{i=0}{C^{k}_{i}r^{2i}},\;\;\;\;\;k\geq
1.\label{13}
\end{equation}
With definition of residues, this expansion permits us to express
the quantization condition (\ref{10}) explicitly in terms of the
coefficients $C_i^k$ as
\begin{equation}
C^{k}_{k-1}= \left(2 n + l + 1 \right) \delta_{1, k}.\label{14}
\end{equation}

It is this quantization condition that makes possible the common 
consideration of the ground and excited states
and permits us to derive the simple recurrent formulae.

The substitution of the series (\ref{12}) and (\ref{13}) into the
system (\ref{9}) in the case $i \neq k-1$ yields the recursion
relation for obtaining the Laurent coefficients 
\begin{eqnarray}
C^{k}_{i}=-{\frac1{2C^{0}_{0}}} \left[(3-2k+2i) C^{k-1}_{i
}+\sum_{j=1}^{k-1}\sum_{p=0}^{i}
C^{j}_{p}C^{k-j}_{i-p} \right. \nonumber \\
\left.
+2\sum_{p=1}^{i}C^{0}_{p}C^{k}_{i-p}+l(l+1)\delta_{2,k}\delta_{0,i}\right].\label{15}
\end{eqnarray}
If $i=k-1$, by equating the explicit expression for $C^k_{k-1}$ to
the quantization condition (\ref{14}) we arrive at the recursion
formulae for the energy eigenvalues
\begin{equation}
2mE_{k}=- C^{k-1}_{k-1}- \sum_{j=0}^{k}\sum_{p=0}^{k-1}
C^{j}_{p}C^{k-j}_{k-1-p} \; .\label{16}
\end{equation}

Thus, the problem of obtaining the energy eigenvalues and eigenfunctions
for the bound-state problem for anharmonic oscillator can be considered
solved. The equations (\ref{15}) and (\ref{16}) have the same simple
form both for ground and excited states and define a useful procedure of
the successive calculation of higher orders of the logarithmic
perturbation theory.

\section{Perturbation corrections to energy eigenvalues}

On applying the recursion relations obtained, analytical
expressions for first corrections to the energy eigenvalues of
the sextic doubly anharmonic oscillator (\ref{0}) are found to be
equal to:
\begin{eqnarray}
E_1  & = &  \frac{1 + 2\,N}{2} \,\omega, \CR\fl E_2 & = &  \frac{\left(
            3 - 2\,L + 6\,\eta  \right) \,{\lambda}}{4\,{\omega }^2},
\CR\fl E_3  & = &  \frac{1 + 2\,N}{8\,{\omega }^5} \,\left(
\left( -21 + 9\,L -
                  17\,\eta  \right) \,{{\lambda}}^2 +
            \left(
                15 - 6\,L +
                  10\,\eta  \right) \,{\omega }^2\,{\mu} \right),
\CR\fl  E_4 & = & \frac{1}{16\,{\omega }^8}\bigg(\left(
            333 + 11\,L^2 - 3\,L\,\left( 67 + 86\,\eta  \right)  +
              3\,\eta \,\left(
                347 + 125\,\eta  \right)  \right) \,{{\lambda}}^3 \CR  &
                -&
        6\,\left(
            60 + 3\,\left( -13 + L \right) \,L + 175\,\eta  - 42\,L\,\eta  +
              55\,{\eta }^2 \right) \,{\omega }^2\,{\lambda}\,{\mu}\bigg),  \label{17} \\
      E_5 & = & - \frac{1 + 2\,N
}{128\,{\omega }^{11}}\CR\fl
      &\times&\bigg(\left(
                    30885 + 909\,L^2 -
                      27\,L\,\left( 613 + 330\,\eta  \right)  + \eta \,\left(
                        49927 + 10689\,\eta  \right)  \right) \,{{\lambda}}^4
                        \CR & - &
                4\left(
                    11220 + 393L^2 -
                      6L\left( 1011 + 475\eta  \right)  + \eta \left(
                        16342 +
                          3129\eta  \right)  \right) {\omega \
}^2{\lambda}^2{\mu}\CR &
                 + &
                2\,\left(
                        3495 + 138\,L^2 + 4538\,\eta  + 786\,{\eta }^2 -
                          30\,L\,\left(
                            63 + 26\,\eta  \right)  \right) \,{\omega }^4\,{{\mu}}^2 \bigg),\nonumber
\end{eqnarray}
where $N = 2 \; n+ l + 1 $, $\eta = N \left(N + 1 \right)$, $L = l
(l+1)$, $m=1$.

 As it is seen, that obtained expansion is the expansion in powers of the
coupling constants. It is also evident that for the energy eigenvalues, 
when $k=1$, we
readily have the oscillator approximation \cite{b20}
\begin{equation}
E_1 = \left( 2 n + l + \frac{3}{2} \right) \omega.
\end{equation}

As a check of the obtained formulae we calculate the energy eigenvalue 
for the ground state of the anharmonic potential

\begin{equation}\label{100}
V(r)=a\; r^2+\frac{b}{3}\; r^4+\frac{c}{9}\; r^6
\end{equation}
admitted the quasi-exact solution \cite{b11}. In this case the expression 
for first corrections to the energy eigenvalue with:
$n=0,\,l=1$ take the form:
\begin{eqnarray}\label{101}
E_1&=&\frac{5\,{\sqrt{a}}}{{\sqrt{2}}}\;,\nonumber \\
E_2&=&\frac{35\,b}{24\,a}\;,\nonumber \\
E_3&=&\frac{35\,\left( -5\,b^2 + 6\,a\,c \right) }
  {96\,{\sqrt{2a^5}}}\;,\nonumber \\
E_4&=&\frac{35\,\left( 475\,b^3 - 864\,a\,b\,c \right) }
  {6912\,a^4}\;, \\
E_5&=&\frac{-35\,\left( 27565\,b^4 - 67488\,a\,b^2\,c +
      17688\,a^2\,c^2 \right) }{110592\,{\sqrt{2a^{11}}}
    }\;,\nonumber \\
E_6&=&\frac{35\,\left( 1451815\,b^5 - 4482360\,a\,b^3\,c +
      2489328\,a^2\,b\,c^2 \right) }{2654208\,a^7}\;.\nonumber
\end{eqnarray}
Under the constraints
\begin{equation}\label{102}
b=\sqrt{2c}\;(2a+\frac{7}{3}\sqrt{2c})^{1/2}
\end{equation}
we arrive at the analytical expression (with $\hbar =  m = 1$)
\begin{equation}\label{103}
E_{0,1}=\frac{5}{2}(2a+\frac{7}{3}\sqrt{2c})^{1/2}
\end{equation}
derived with the supersymmetric consideration in \cite{b11}.

Just as it was expected, the obtained series (\ref{17}) 
for the sextic doubly anharmonic oscillator is divergent 
that compels us to resort to modern procedures of
summation of divergent series. One of the most common among them
are various versions of the renormalization procedures intended to
 reorganize a given series into another one with better convergence
properties.

It was found that the proposed technique is easily adapted to 
any renormalization scheme.
Now we consider the case of the renormalization of a frequency. For this
purpose it is enough to think of the harmonic oscillator frequency
incoming in the potential as a function of Plank's constant variable,
with subsequent its expansion in an $\hbar$-series. However, for later
use it is more convenient to take
\begin{eqnarray}{\label{1.33}}
\omega^2=\sum\limits_{k=0}^{\infty}\omega_k^2 \hbar^k .
\end{eqnarray}
The relations for the coefficients $ C_{i}^{k}$ then reads
\begin{eqnarray}{\label{xx}}
C^{0}_{0}&=&-m\omega_{0} ,\; \;C^{0}_{i}={\frac1{2m\omega_{0}}}
({\sum_{p=1}^{i-1}{C^{0}_{p} C^{0}_{i-p}-2mv_{i}}}), \;i\geq 1,\nonumber \\
C^{k}_{i} &=& -{\frac1{2C^{0}_{0}}} \Bigg[ {-
m^{2}\omega^{2}_{\,k}\delta_{i,k}+(3-2k+2i) C^{k-1}_{i } }
 \\ && +\sum_{j=1}^{k-1}\sum_{p=0}^{i}
C^{j}_{p}C^{k-j}_{i-p}
+2\sum_{p=1}^{i}C^{0}_{p}C^{k}_{i-p}+l(l+1)\delta_{2,k}\delta_{0,i}\Bigg],\;k\geq
1,\nonumber
\end{eqnarray}
but the recursion system for the energy eigenvalues does not change.

The coefficients $\omega_k^2$ are defined by the chosen version of
renormalization. In practice, the one-parameter schemes are
usually used. They are obtained with the truncation of the series (\ref{1.33}) 
as
\begin{eqnarray}{\label{1.37}}
\omega^2=\omega_0^2+\omega_k^2 \hbar^k ,
\end{eqnarray}
where  $\omega_0$ is a trial frequency and an order in $\hbar$
of the remainder is determined by the anharmonic term in a
potential. In the case of our potential (\ref{0}), taking into
account the order in $\hbar$ of the dimensionless parameter
of the expansion, we must put $k=1$.

It has been recognized that for three-dimensional sextic 
doubly anharmonic oscillator the schemes of both the minimal 
difference, $E_N(\omega_0)=0$, and the minimal sensitivity,
$(d/d\omega_0)E_N(\omega_0)=0$,\cite{b5} appear to give 
approximately an equal accuracy in vast range of values of
the anharmonic parameters.

Typical results of the calculation are presented in the Table
where the sequences of the partial sums of $N$
renormalized corrections to the energy eigenvalues of the
spherical anharmonic oscillator with the potential
$V(r)=\frac{1}{2}r^2 +\lambda r^4+ \mu r^6 $ is compared
with the results, $E_\text{num}$, obtained by the numerical
integration for the values $\lambda=\mu$, 
($\hbar = m =\omega= 1$). The free parameter $\omega_0$ 
in the one-parameter scheme (\ref{1.37})  is defined order
by order from the equation of minimal
sensitivity, $(d/d\omega_0) E_N (\omega_0) = 0$, with choosing
such the root that obeys the condition of the flattest extremum 
\cite{b21}.

From the Table it is seen that the one-parameter renormalization 
procedure gives the accuracy which is quite sufficient for 
practical purposes.

\begin{table}[ht]\label{table1}
\begin{center}
\caption{ The sequences of the partial sums of $N$
renormalized corrections. } \vspace{10pt}
\begin{tabular}{c|ll|ll|ll}
\hline \vphantom{\Big(} $N$ & \hss $n=0, \;l=0$ & & \hss $ n=1,\;
l=0$&
&\hss $ n=1,\; l=1$ &\\
\cline{2-7} \vphantom{\Big(}   & $\lambda=0.01$ & $\lambda=10$ &
$\lambda=0.01$
  & $\lambda=10$ & $\lambda=0.01$ & $\lambda=10$ \\
\hline
2 & 1.535791 & 5.382133 & 3.670797 & 15.99931 & 4.765971 & 22.01213 \\
5 & 1.616383 & 6.042488 & 4.144668 & 18.75063 & 5.545735 & 26.18935 \\
10 & 1.620603 & 6.097330 & 4.184985 & 19.07378 & 5.614234 & 26.67913 \\
15 & 1.621682 & 6.125723 & 4.193532 & 19.15066 & 5.629195 & 26.79595 \\
20 & 1.621688 & 6.126339 & 4.223470 & 19.57157 & 5.683960 & 26.84757 \\
25 & 1.621689 & 6.126361 & 4.223784 & 19.57557 & 5.684735 & 26.87657 \\
30 & 1.621690 & 6.126367 & 4.223822 & 19.57838 & 5.685455 & 27.40164 \\
35 & 1.621690 & 6.126369 & 4.223840 & 19.57880 & 5.685520 & 27.39683 \\
40 & 1.621690 & 6.126370 & 4.223842 & 19.57928 & 5.685545 & 27.39740 \\
\hline
$E_\text{num}$ & 1.621690 & 6.126371 & 4.223843 & 19.57939 & 5.685575 & 27.39812\\
\hline
\end{tabular}
\end{center}
\end{table}
In conclusion, a new useful technique for deriving results of the
logarithmic perturbation theory has been developed. Based upon the
$\hbar$-expansions and suitable quantization conditions, new handy
recursion relations for solving the bound-state problem for a spherical
anharmonic oscillator have been obtained. These relations can be applied
to excited states exactly in the same manner as to the ground states
providing, in principle, the calculation of the perturbation corrections
of large orders in the analytic or numerical form. Besides this
remarkable advantage over the standard approach to the logarithmic
perturbation theory, our method does not imply knowledge of the exact
solution for zero approximation, which is obtained automatically. And at
last, the recursion formulae at hand are easily adapted to the use of any
renormalization scheme for improving the convergence of obtained series.

The proposed technique has been applied for investigation of the 
bound-state problem of the three-dimensional sextic doubly
anharmonic oscillator. It has been observed that the perturbation
series of the logarithmic perturbation theory are divergent
but the use of the one-parameter renormalization 
procedure gives the accuracy which is quite sufficient for 
practical purposes.

\begin{acknowledgments}
This research was supported by a grant N~0103U000539 from the Ministry
of Education and Science of Ukraine which is gratefully acknowledged.
\end{acknowledgments}

\end{document}